\documentclass{PoS}

\usepackage{epsfig,macros_static}

\newcommand{\ftext}[1]{\fbox{ {#1} }}
\newcommand{\onecol}[2]{
        \begin{minipage}[t]{#1}{#2\vfill} \end{minipage}
        }

\PoS{PoS(LAT2005)223}

\title{A strategy for the computation of $m_b$ including $1/m$ terms }

\ShortTitle{A strategy for the computation of $m_b$ including $1/m$ terms }

\author{\epsfxsize=2.5 true cm
	\epsfbox{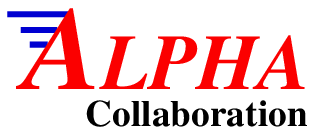} \vspace{-1.8cm} \hfill
{\onecol{2.8cm}{\vspace{-1cm} \it DESY 05-170 \\ SFB/CPP-05-58}}
}

\author{Michele Della Morte \thanks{
This work has been supported by the Deutsche
Forschungsgemeinschaft (DFG) in
SFB Transregio 9 ``Computational Particle Physics''.}\\

        Institut f\"ur Physik, Humboldt Universit\"at,
        Newtonstr. 15, 12489 Berlin, Germany\\
        E-mail: \email{dellamor@physik.hu-berlin.de}}

\author{Nicolas Garron, \speaker{Rainer Sommer}
        \\
        DESY, 
        Platanenallee 6, 
        15738 Zeuthen, 
        Germany
        \\
        E-mail: \email{Nicolas.Garron@desy.de, Rainer.Sommer@desy.de}}

\author{Mauro Papinutto \\
        NIC/DESY, 
        Platanenallee 6, 
        15738 Zeuthen, 
        Germany
        \\
         E-mail: \email{Mauro.Papinutto@desy.de}}

\abstract{We consider HQET including the first order
correction in $1/m$. A strategy for the computation of 
the b-quark mass following the scheme

\newcommand{\gre}{\cgre}
\newcommand{\blu}{\cblu}
\newcommand{\asylat}[1]  % \unitlength 1mm for 8 cm lattice!
{\begin{picture}(160,160)
\unitlength #1
\linethickness{0.1mm}
\blu\multiput(0,0)(0,10.0){8}{\line( 1, 0){80.0}}
\blu\multiput(0,0)(2.5,0){32}{\line( 0, 1){80.0}}
\end{picture}
}
\newcommand{\vbiglat}[1]  % \unitlength 1mm for 16 cm lattice!
{\begin{picture}(160,160)
\unitlength #1
\linethickness{0.1mm}
\blu\multiput(0,0)(0,10.0){16}{\line( 1, 0){160.0}}
\blu\multiput(0,0)(10.0,0){16}{\line( 0, 1){160.0}}
\end{picture}
}
\newcommand{\vvbiglat}[1]  % \unitlength 1mm for 32 cm lattice!
{\begin{picture}(320,280)
\unitlength #1
\linethickness{0.1mm}
\blu\multiput(0,0)(0,10.0){28}{\line( 1, 0){320.0}}
\blu\multiput(0,0)(10.0,0){32}{\line( 0, 1){280.0}}
\end{picture} 
}

\newcommand{\bwave}[1]  % \unitlength 1mm for a=1 cm lattice!
{\unitlength #1
\begin{picture}(50,20)
\linethickness{0.1 mm}
\mgt \qbezier(0,0)(1.2,20)(1.9,20)     % 0,pi/2 (pi=120)
     \qbezier(1.9,20)(2.5,20)(3.75,0)     % pi/2,pi
      \qbezier(3.75,0)(4.95,-20)(5.6,-20)
      \qbezier(5.6,-20)(6.3,-20)(7.5,0)
\end{picture}
}
\newcommand{\bwavelnod}[1]  % \unitlength 1mm for a=1 cm lattice!
{\unitlength #1
\begin{picture}(50,20)
\linethickness{0.7 mm}
\mgt \qbezier(0,0)(24,20)(37.5,20)     % 0,pi/2 (pi=120)
     \qbezier(37.5,20)(50,20)(75,0)     % pi/2,pi
      \qbezier(75,0)(99,-20)(112.5,-20)
      \qbezier(112.5,-20)(126,-20)(150,0)
\end{picture}
}
\newcommand{\bwaveasy}[1]  % \unitlength 1mm for a=1 cm lattice!
{\unitlength #1
\begin{picture}(50,20)
\linethickness{0.7 mm}
\mgt \qbezier(0,0)(12,20)(19,20)     % 0,pi/2 (pi=120)
     \qbezier(19,20)(25,20)(37.5,0)     % pi/2,pi
      \qbezier(37.5,0)(50,-20)(56,-20)
      \qbezier(56,-20)(69,-20)(75,0)
\end{picture}
}

\newcommand{\piwave}[1]  % \unitlength 1mm for a=1 cm lattice!
{\unitlength #1
\begin{picture}(50,20)
\linethickness{0.2 mm}
\gre \qbezier(0,0)(40,20)(70,20)      % 0,pi/2 (pi=140)
     \qbezier(70,20)(100,20)(140,0)   % pi/2,pi
      \qbezier(140,0)(170,-20)(210,-20)
      \qbezier(210,-20)(240,-20)(280,0)
\end{picture}
}
\newcommand{\pihwave}[1]  % \unitlength 1mm for a=1 cm lattice!
{\unitlength #1
\begin{picture}(50,20)
\linethickness{0.7 mm}
\gre \qbezier(0,0)(40,20)(70,20)      % 0,pi/2 (pi=140)
     \qbezier(70,20)(100,20)(140,0)   % pi/2,pi
\end{picture}
}

\vspace{2.5cm}\hspace{1.5cm}
\begin{picture}(8,20)(0,0)
\small
  \unitlength 0.4cm
  \put(2,6){\ftext{experiment}}            \put(16.5,6){\ftext{Lattice with 
$a\mq\ll 1$}} 
  \put(2,4){ $\mB=5.4\,\GeV$}    \put(16,4){ $\Phi_1(L_1,M),\Phi_2(L_1,M)$} 
  \linethickness{0.3mm}\put(5.3,3.5){\vector(0,-1){1.5}}
  \linethickness{0.3mm}\put(17,3.5){\vector(0,-1){1.5}}
  \linethickness{0.3mm}
  \put(0,0.5){ $\Phi_1^\mrm{HQET}(L_2),\Phi_2^\mrm{HQET}(L_2)$}
  \put(16,0.5){ $\Phi_1^\mrm{HQET}(L_1),\Phi_2^\mrm{HQET}(L_1)$}
  \put(14.7,0.7){\vector(-1,0){5.0}}
  \put(10.5,1.1){$\sigmam(u_1)$}
  \put(9.5,-0.4){$\sigmakin_1(u_1),\sigmakin_2(u_1)$}
  \put(10,3.0){\small $L_2 = 2 L_1$}
\end{picture}
\vspace{0.3cm}
is discussed. Only two quantities $\Phi_{1/2}$ 
have to be considered in order to match QCD and HQET,
since the spin-dependent interaction is
easily eliminated due to the spin symmetry of the 
static theory. Quite simple formulae relate the 
renormalization group invariant b-quark mass ($\Mb$) to
the B-meson mass. All entries in these formulae
are non-perturbatively defined and can be computed
in the continuum limit of the lattice regularized 
theory. For the numerically most critical part, we illustrate the
cancellation of power divergences by a numerical example.\\
Numerical results for the $1/m$ correction to
$\Mb$, are presented in a companion talk.

}

\FullConference{XXIIIrd International Symposium on Lattice Field Theory\\

                 25-30 July 2005\\

                 Trinity College, Dublin, Ireland}

\begin{document}
%%%%%%%%%%%%%%%%%%%%%%%%%%%%%%%%%%%%%%%%%%%%%%%%%%%%%%%%%%%%%%%%%%%%%%%%%%%%%%%
\section{Introduction}

Although HQET is  the most natural effective theory for heavy-light
systems, its lattice regularized version has practically 
only been used at lowest order.
Indeed, a strategy to overcome the
problem of power divergent mixings~\cite{stat:MaMaSa}, was only
found rather recently \cite{hqet:pap1}. Its potential
was demstrated by a computation of the b-quark mass to lowest 
non-trivial order in $1/m$, the static approximation.
Here we fill the formalism of \cite{hqet:pap1}, sketched
in the abstract, with practicable definitions in terms
of \SF correlation functions and give a concrete formula for the
$1/m$-correction to the quark mass. 

Neglecting $1/m^2$ corrections -- as throughout this report -- we write the HQET Lagrangian
\bes
\lag{HQET} &=&  \lag{stat}(x) - \omegaspin\Ospin(x) 
        - \omegakin\Okin(x)\\
 && \Ospin = \bar\heavy{\bf\sigma B}\heavy\,,\quad 
    \Okin  = \bar\heavy{\bf D}^2\heavy \,,
\ees
such that the classical values for the coefficients
are $\omegakin=\omegaspin=1/(2m)$. Since expectation values
\bes
  \langle \op{} \rangle &=& 
    \langle  \op{}  \rangle_\mrm{stat} 
        +   \omegakin \langle  \op{}  \rangle_\mrm{kin} 
        +  \omegaspin \langle  \op{}  \rangle_\mrm{spin} \,,
        \label{e:expa}\\
        &&  \langle  \op{}  \rangle_\mrm{kin} = 
         \sum_x \langle  \op{} \Okin(x) \rangle_\mrm{stat}\,,\quad
            \langle  \op{}  \rangle_\mrm{spin} = 
              \sum_x \langle \op{} \Ospin(x) \rangle_\mrm{stat} 
        \label{e:exp}
\ees
are defined through {\em insertions} of the higher dimensional terms 
$\Okin, \Ospin$ {\em in the static theory}, they are renormalizable by power counting. 
However,
in order to have a well defined continuum limit 
% (or definition beyond perturbation theory) 
the bare, dimensionful, couplings 
$\omegakin,\omegaspin$ have to be determined 
non-perturbatively~\cite{stat:MaMaSa,hqet:pap1}. In the framework of lattice
QCD, this is possible by matching a number of observables,
$\Phi_i, \; i=1\ldots n,$
between QCD and HQET, thus retaining the predicitivity of QCD. 
It is essential to note that this matching can be carried out
in a finite volume of linear extent $L_1 \simeq 0.4\,\fm$, where
heavy quarks can be simulated with a relativistic 
action~\cite{hqet:pap1,hqet:pap2,hqet:pap3}. 

Since the lowest order theory is spin-symmetric, it is trivial
to form spin-averages which are independent of $\omegaspin$. One
thus expects that $n=2$ is sufficient for a computation of the
quark mass (in addition to $\omegakin$ there is an overall 
(state-independent) shift of energy levels, which we denote
by $\mhbare$). For unexplained notation we refer to \cite{hqet:pap1}.

% \input lattices.tex
% \input f_mbstratfirst.tex

%%%%%%%%%%%%%%%%%%%%%%%%%%%%%%%%%%%%%%%%%%%%%%%%%%%%%%%%%%%%%%%%%%%%%%%%%%%%%%%
\section{Basic observables}

We consider the spin-symmetric combination
\bes
  \foneav(\theta,T) &=& \zzeta^4 \left\{\fone(\gamma_5)\right\}^{1/4}  
                \left\{\fone(\gamma_1)\right\}^{3/4}\,, 
\ees
formed from the boundary to boundary correlation functions
\bes
  \fone(\Gamma) &=& 
  -{a^{12} \over 2L^6}\sum_{\vecu,\vecv,\vecy,\vecz}
  \left\langle
  \zetalbprime(\vecu)\Gamma\zzetaprime_{\rm b}(\vecv)\,
  \zetabar_{\rm b}(\vecy)\Gamma\zetal(\vecz)
  \right\rangle\,,
\ees
of the QCD \SF of size $T\times L^3$ and a periodicity
phase $\theta$ \cite{pert:1loop} for the quark fields. 
Replacing the b-quark field by the effective field 
$\heavy$, using eq.(\ref{e:expa},\ref{e:exp}), and accounting for the 
multiplicative renormalization of the boundary quark
fields $  \zeta\,,\;\zetabar$ one finds the $1/m$ expansion
\bes
  \foneav &=& \zzetah^2\zzeta^2 \rme^{-\mhbare T} 
        \left\{ \fonestat + \omegakin \fonekin
        \right\}\,, 
\ees
where the aformentioned energy shift $\mhbare$ enters.
Deviating from the choice in \cite{hqet:pap1}, we now define\footnote{
In the static computation of \cite{hqet:pap1} the logarithmic 
derivative $\meff$ of the correlation function $\fa$ of the axial current
with a boundary operator was used as a quantity to match effective 
theory and QCD. Including $1/m$ terms its expansion reads 
\bes
  \fa &=& \zahqet \zzetah\zzeta \rme^{-\mhbare x_0} 
        \left\{ \fastat + \cahqet \fdeltaastat + \omegakin \fakin
                + \omegaspin \faspin
        \right\}\,, \label{e:faexp} 
\ees
with the term $\fdeltaastat$ due to the $1/m$ correction to the static
axial current. While $\omegaspin$ represents no problem, an extra 
observable is needed to fix $\fdeltaastat$. Here, we avoid this complication
by working exclusively with $\foneav$.
}
\bes
  \Phi_1(L,M) &=& \ln \left(\foneav(\theta,T) / \foneav(\theta',T)\right) 
                  - \ln \left(\fonestat(\theta,T) / \fonestat(\theta',T)\right) \\[-1.2ex]
        && \qquad \qquad\qquad\qquad\qquad \qquad\qquad\qquad\qquad\qquad\qquad\quad
        \mbox{for} \quad  T=L/2 \,, \nonumber \\[-1.2ex]
  \Phi_2(L,M) &=& \frac{L}{2a} \ln \left(\foneav(\theta,T-a)/\foneav(\theta,T+a)
        \right)\,,       
\ees
with the expansion
\bes
  \Phi_1(L,M) &=& \omegakin \ratonekin\,,  \label{e:phi12}
        \quad
  \Phi_2(L,M) = L \, \left(\mhbare + \meffstat_1+\omegakin \meffkin_1\right)\, 
         \\
  \ratonekin &=& {\fonekin(\theta,T)\over\fonestat(\theta,T)}
                           - {\fonekin(\theta',T) \over \fonestat(\theta',T)}
                       \,, 
                        \\
  \meffstat_1 &=& \frac1{2a} \ln 
        \left(\fonestat(\theta,T-a)/\fonestat(\theta,T+a)
        \right)\,,
        \\ 
  \meffkin_1 &=& \frac1{2a}\left({\fonekin(\theta,T-a)\over\fonestat(\theta,T-a)}
                           - {\fonekin(\theta,T+a) \over \fonestat(\theta,T+a)}
                        \right)\,.
\ees

%%%%%%%%%%%%%%%%%%%%%%%%%%%%%%%%%%%%%%%%%%%%%%%%%%%%%%%%%%%%%%%%%%%%%%%%%%%%%%%%

\section{Step scaling functions}

We choose $L_1\approx 0.4\fm$, where a computation of $\Phi_i(L_1,\Mbeauty)$ 
is possible in lattice QCD (while at significantly larger values, $L_1/a$ 
would have to be
too large in order to control $a^2$ effects). 
From \eq{e:phi12} one then gets 
$\omegakin,\mhbare$ for lattice spacings $a=\frac{a}{L_1} \times 0.4 \fm$.
On the other hand, contact to physical observables, e.g. the B-meson mass
is made in large volume, where finite size effects are exponentially
small.
For reasonable
values $a/L_1 = 1/12$ and $L_\infty \simeq 1.5 \fm$ at the same lattice spacing, 
one needs $L_\infty/a \sim 50$. This situation is avoided
by first computing step scaling functions
which connect $\Phi_i(L_1,M)$ to  $\Phi_i(L_2,M), L_2=2L_1$ and then connecting 
to large volume. 

With the \SF coupling, $u=\gbar^2(L)$,  everywhere, the 
continuum step scaling functions $\sigma$
are defined by
\bes
   \Phi_1(2L,M) = \sigmakin_1(u)  \Phi_1(L,M)\,,\quad
   \sigmakin_1(u) = \lim_{a/L \to 0} \left.{\ratonekin(2L) \over  
              \ratonekin(L) }\right|_{u=\gbar^2(L)}\,
\ees
and
\bes
   \label{e:sigma2kin}
   \Phi_2(2L,M) &-& 2\Phi_2(L,M)= \sigmam(u) +
            \left[\omegakin\, 2L\, (\meffkin_1(2L) - \meffkin_1(L))\right] \\
            &=&  \sigmam(u) +  \sigmakin_2(u)\, \Phi_1(L,M)\,,
            \quad   \sigmakin_2(u) =  
                \lim_{a/L \to 0}  2L\left. {\meffkin_1(2L) - \meffkin_1(L)
                                        \over
                                  \ratonekin(L)  }\right|_{u=\gbar^2(L)} \,.
	    \nonumber
\ees
Here the static step scaling function
\bes
  \sigmam(u) =  
                \lim_{a/L \to 0}  2L\,\left[ \meffstat_1(2L) - \meffstat_1(L)
                                \right]_{u=\gbar^2(L)} \,,
\ees
is not identical to $\sigmam (u)$ defined earlier~\cite{hqet:pap1},
since $\meffstat_1$ differs from $\meffstat$ defined there. Note that the 
step scaling functions are independent of $M$, but $\Phi_i(L,M)$ have a
mass dependence from fixing $\Phi_i(L_1,M)$ in the full theory.

%%%%%%%%%%%%%%%%%%%%%%%%%%%%%%%%%%%%%%%%%%%%%%%%%%%%%%%%%%%%%%%%%%%%%%%%%%%%%%%%
\section{Large volume}

The connection of $\Phi_i$ to the spin-averaged B-meson mass, $\mb$, is
\bes
\label{e:large0}
 L \mb &-& \Phi_2(L,M) =  \left[L \,(E^\mrm{stat} - \meffstat_1(L))\right] 
        +  \left[ L \,\omegakin \,(\hat{E}^\mrm{kin} - \meffkin_1(L))\right]  \\
        \label{e:largel}
        &=&  \left[ L \,(E^\mrm{stat} - \meffstat_1(L)) \right]
        + \rho(u) \Phi_1(L,M)\,,\quad
        \rho(u) =  \lim_{a/L \to 0} 
        L\left.{\hat{E}^\mrm{kin} - \meffkin_1(L) \over \ratonekin(L)
        }\right|_{u=\gbar^2(L)} \,. \nonumber
\ees
Here we have used the abbreviations
\bes
  E^\mrm{stat} = \lim_{L\to \infty} \meffstat_1(L) \,,\qquad
  \hat{E}^\mrm{kin}  = \lim_{L\to \infty} \meffkin_1(L)\,, 
\ees
where $ E^\mrm{stat}$ is the (unrenormalized) energy in large volume 
in the spin-averaged
B-channel in static approximation and $\omegakin\,\hat{E}^\mrm{kin}$ 
is its $1/m$ correction. The hat on $\hat{E}^\mrm{kin}$ is to 
remind us that this quantity turns into
an energy only upon multiplication with the dimensionful 
$\omegakin$. Its numerical evaluation has already been
investigated in \cite{lat04:stephan}. We use $[...]$ braces to indicate 
combinations which have a continuum limit by themselves. For example,
the two terms in \eq{e:large0} can be computed with different 
regularizations  if this is useful.

\section{Final equation}

The above equations are now easily combined to yield 
the $1/m$ 
correction, $\mb^{(1)}$, to the (spin-averaged) B-meson mass via
($L_2=2L_1$),
%$L_1 \approx 0.4\,\fm$)
\bes
 \mb &=& \mb^\mrm{stat} +  \mb^{(1)} =  \mb^\mrm{stat} +  \mb^{(1a)}+\mb^{(1b)}\,,\\[1ex]
   L_2\,\mB^\mrm{stat}(M) 
        &=& \left[L_2\,(E^\mrm{stat} - \meffstat_1(L_2))\right] 
        + \sigmam(u_1) + 2\,\Phi_2(L_1,M)\\[0ex]
 L_2\,\mb^{(1a)}(M) &=& \sigmakin_2(u_1) \,\Phi_1(L_1,M)\,, 
        \qquad        \; u_i=\gbar^2(L_i) \\
 L_2\,\mb^{(1b)}(M) &=&  \left [L_2\,(\hat{E}^\mrm{kin} - \meffkin_1(L_2)) 
                         \omegakin\right] 
       =  \rho(u_2)\, \sigmakin_1(u_1) \Phi_1(L_1,M) 
         \,. \nonumber
  \label{e:master}
\ees
Again, 
terms in braces have a continuum limit. While $ \mb^{(1a)}$ is purely derived
from finite volume, the term $\mb^{(1b)}$ involves a large volume
computation. 

Starting from $\Mb^\mrm{stat}$, the solution of the leading order equation,
\bes
 \mB^\mrm{exp} &=& \mB^\mrm{stat}(\Mb^\mrm{stat})\,,
\ees
\vspace{-0.3cm}
and the slope
\vspace{-0.3cm}
\bes
 S&=& \left.{\mrm{d} \over \mrm{d} M}   \mB^\mrm{stat} \right|_{M= \Mb^\mrm{stat}}
     = {1 \over L_1} \left.{\mrm{d}  \over \mrm{d}M} \Phi_2(L_1,M)\right|_{M= \Mb^\mrm{stat}}\,,
\ees
we finally obtain the first order correction
$\Mb^{(1)}$ to the RGI b-quark mass
\bes 
   \Mb &=&  \Mb^\mrm{stat} + \Mb^{(1)}\,, \quad
 \Mb^{(1)} =  - {1\over S}  \mb^{(1)} \,.
\ees
The final  uncertainty 
for $\Mb$ {\em due to the $1/m$ expansion} 
is of order $\rmO(\Lambda_{\rm QCD}^3/\Mb^2)$,
which translates into a numerical estimate of $\MeV$ scale.
It is thus clear that other sources of error will dominate
in a practical calculation.
Note that the precise value for $\mb$ matters. 
One should use the spin-averaged mass
%\bes
$
  \mb^\mrm{experimental} = \frac14 m_{\rm B^0} + \frac34 m_{\rm B^*_0} = 
        [\frac14\, 5279 +   \frac34 \, 5325] \, \MeV = 5314\, \MeV
$
%\ees
if one can extrapolate $E$ to the chiral limit of the light quark or 
\bes
  \mb^\mrm{experimental} &=& 
        m_{\rm B_s} + \frac34 m_{\rm B^*_0} - \frac34 m_{\rm B_0} %\\
        =%&=& 
        [ 5370 +   \frac34 (\, 5325 -  \, 5279)] \, \MeV = 5405\, \MeV
\ees
if one works directly with a strange quark (as light quark).
The latter formula neglects the dependence of the spin splitting on
the light quark mass. 

%%%%%%%%%%%%%%%%%%%%%%%%%%%%%%%%%%%%%%%%%%%%%%%%%%%%%%%%%%%%%%%%%%%%%%%%%%%%%%

\section{Remarks}

The following facts are worth noting. 
\begin{itemize}
\vspace{-0.15cm}
\item The $1/m$ expansion in heavy light systems is an expansion
      in terms of $\Lambda_\mrm{QCD}/m$, where all external scales
      have to be of order $\Lambda_\mrm{QCD}$. This applies in particular
      to our scale $L_1^{-1}$. Indeed, numerically it is rather close to
      $\Lambda_\mrm{QCD}$ and explicit investigations
	\cite{hqet:pap2,hqet:pap3} have shown that
      the $1/m$-expansion is well behaved even when $L^{-1}$ is a factor
      two larger. 
\vspace{-0.15cm}
\item In our static computation \cite{hqet:pap1,hqet:pap2}, we made the
      more natural choice $\meff$
      instead of $\Gamma_1$. 
      Although it is
      advantageous to use $\Gamma_1$ when one includes the $1/m$ terms,
      the strategy can easily be formulated with $\meff$, at the expense
      of introducing a third quantity $\Phi_i$ to fix $\cahqet$. Since this 
	will certainly
      be required for the computation of the $1/m$-correction to $\fb$, we will
      follow also that approach. 
\vspace{-0.15cm}
\item Note that at each order $k$  in the expansion, the result is ambiguous
      by terms of order $1/m^{k+1}$. Thus both $\Mb^{(1)}$ and $\Mb^\mrm{stat}$
      have an order $1/m$ ambiguity (e.g. they change when $L_1$ is changed), while in their sum
      $ \Mb =  \Mb^\mrm{stat} + \Mb^{(1)}$ the ambiguity is reduced to 
      $1/m^2$.
\vspace{-0.15cm}
\item In the present formulation of the effective theory, the
      $1/m$-terms approach the continuum with an asymptotic rate $\propto a$,
      in contrast to the leading order terms where this is 
	$\propto a^2$~\cite{hqet:pap1}.
\vspace{-0.15cm}
\item Let us comment just on one numerical result at that point. 
	The computation of $\sigmakin_2(u_1)$,  \eq{e:sigma2kin}, 
	involves the difference of $\meffkin_1(2L) - \meffkin_1(L)$,
	where power divergent contributions cancel. As a typical case
	we choose $L/a=12\,,\;T/a=6$,
	and the static action HYP2 (see \cite{stat:actpaper}), where
	our simulations yield
	$a^2\meffkin_1(2L)=0.5631(6)\,,\;\;a^2\meffkin_1(L)=0.5595(2)$,
	demonstrating a considerable cancellation.
      A detailed account of numerical results is presented in \cite{lat05:nicolas}.
\end{itemize}

\bibliography{refs}           %or whatever your .bib file is
\bibliographystyle{h-elsevier}

%\begin{thebibliography}{99}

%  \bibitem{...} ....
%  \bibitem{...} ....

%\end{thebibliography}

\end{document}